\documentstyle[12pt]{article}
\let\saveheight\textheight
\let\savewidth\textwidth
\input FEYNMAN.tex
\let\textheight\saveheight
\let\textwidth\savewidth
\def\PM{\hphantom{-}} \def\PV{\hphantom{|}} \def\PZ{\hphantom{0}}
\def\cf{{\em cf.\/}}
\def\eg{{\em e.g.\/}} \def\ie{{\em i.e.\/}} \def\etc{{\em etc.\/}}
\def\cA{c_{\scriptscriptstyle A}} \def\cV{c_{\scriptscriptstyle V}}
\def\gA{g_{\scriptscriptstyle A}} \def\gV{g_{\scriptscriptstyle V}}

\def\SO{\Sigma^0}  \def\SM{\Sigma^-} \def\SP{\Sigma^+} \def\SC{\Sigma^\pm}
\def\LO{\Lambda^0} \def\XM{\Xi^-}    \def\XO{\Xi^O}
\def\Kl3{K_{\ell3}}
\def\tr{\mathop{\mathrm{tr}}\nolimits}
\def\bra#1{\left\langle#1\>\!\right|\>\!}
\def\ket#1{\>\!\left|\>\!#1\right\rangle}
\def\ftnt{\,\footnote}
\def\NoCol{\multicolumn{1}{c}{}}
\makeatletter
\def\prefix#1#2#3{\mathchoice
 {\let\s@vestyle\displaystyle\@prefix{#1}{#2}{#3}}
 {\let\s@vestyle\textstyle\@prefix{#1}{#2}{#3}}
 {\let\s@vestyle\scriptstyle\@prefix{#1}{#2}{#3}}
 {\let\s@vestyle\scriptscriptstyle\@prefix{#1}{#2}{#3}}\mkern-4mu#3}
\def\@prefix#1#2#3{\setbox0\hbox{$\s@vestyle#3$}\setbox2\null\relax
 \ht2\ht0 \dp2\dp0 \box2#1#2}
\newlength{\@bxntwd}
\def\boxnotes#1{\settowidth{\@bxntwd}{#1}\relax
 \begin{minipage}{\@bxntwd} \let\footnoterule=\relax #1 \end{minipage}}
\makeatother
\begin{document}
\def\thefootnote{\fnsymbol{footnote}}
\begin{flushright}
\raisebox{18pt}[0pt][0pt]{\vbox{UNICO-97-010\\hep-ph/9710458}}
\end{flushright}
\begin{center}
{\LARGE\bf\ A New Quark Model Approach\\ to Hyperon Semi-Leptonic Decays\relax
 \footnote{Presented at the Topical Workshop on Deep Inelastic Scattering
 off Polarized Targets: Theory Meets Experiment---Zeuthen, Sept.~1997}}

\vspace{1cm}
{Philip G. Ratcliffe}

\vspace*{1cm}
{\it Istituto di Scienze MM.FF.CC.,\\
     II Facolt\`a di Scienze MM.FF.NN.,\\
     Universit\`a di Milano sede di Como,\\
     via Lucini 3, 22100 Como, Italy}\\[3pt]
and\\[3pt]
{\it INFN -- Sezione di Milano}\\

\vspace*{1.5cm}

\end{center}
\begin{abstract}
A new analysis of hyperon semi-leptonic decay data is presented, based on 
simple and general arguments for an SU(3)-breaking structure of the couplings
involved.
The results obtained are compatible with those of earlier investigations and 
highlight the uncertainties inherent in extraction of the $V_{us}$ 
Cabibbo-Kobayashi-Maskawa matrix element from such data, and the need for 
complete analyses.
For $F$ and $D$, the results can be summarised by the ratio: $F/D=0.57\pm0.01$;
while as far as the value of $V_{us}$ is concerned, the only conclusion to be
drawn is that, insofar as SU(3) breaking is still not totally under control,
the value obtained is {\em compatible\/} but not {\em competitive\/} with that
obtained from kaon semi-leptonic decays.
\\[6pt] PACS: 13.30.Ce, 13.88.+e, 11.30.Hv, 13.60.Hb
\end{abstract}
\section{Introduction}

\vspace{1mm}\noindent
Hyperon semi-leptonic decays (HSD) represent the only presently available 
source for obtaining the two axial-coupling parameters $F$ and $D$ separately 
(the sum being known rather precisely from neutron beta decay alone).
In addition, these data are used for extracting a value of the 
strangeness-changing Cabibbo-Kobayashi-Maskawa (CKM) \cite{CKM} matrix element, 
$V_{us}$.
The first two of the above parameters are vital to the analysis of data on the
longitudinally polarised deep-inelastic scattering (DIS) structure functions,
$g_1^{p,n}(x)$, as they provide the overall normalisation for certain flavour
non-singlet combinations of the different quark-parton helicity-dependent
distributions.
As the precision of such measurements improves, a reliable and accurate
evaluation of $F$ and $D$ (or, more to the point, of the combination orthogonal 
to the sum) becomes ever-more necessary.

The third parameter mentioned above, $V_{us}$, is also obtainable from the 
so-called $\Kl3$ decays~\cite{Leu84a}.
The most recent Particle Data Group (PDG) publication \cite{PDG96a} quotes the
two values:
\begin{eqnarray}
 V_{us}^{\Kl3}         & = & 0.2196   \pm 0.0023\ \cite{Leu84a}, \\
 V_{us}^{\mathrm{HSD}} & = & 0.222 \PZ\pm 0.003\ .\label{eq:VusHSD}
\end{eqnarray}
From these, the PDG obtains the following world average~\cite{PDG96a}:
\begin{eqnarray}
 V_{us}^{\mathrm{ave}} & = & 0.2205 \pm 0.0018\ .\hphantom{[0}
\end{eqnarray}
Note the smallness of the error ascribed to the HSD measurement and the
non-negligible weight with which it therefore contributes to the world average.
Note also that the data used in eq.~(\ref{eq:VusHSD}) come only from the WA2
experiment \cite{SPS} and the inclusion of later data \cite{Hsu88a} would 
clearly alter this number slightly.
Moreover, a different approach to SU(3) breaking in HSD \cite{Gar92a} gives 
slightly higher value than that quoted.
I shall return to these points more in detail in section~\ref{sec:fits}.

The description of HSD in terms of just the three parameters discussed above
relies, of course, on the validity of SU(3) flavour symmetry in hadron 
dynamics.
Now, while isospin breaking is relatively small (of the order of less than a
percent) and the QED radiative corrections are known and may be incorporated
into the analysis where necessary, symmetry breaking is of the order of ten
percent or more in the hypercharge direction.
Inevitably then, in order to extract $F$ and $D$ to a precision of order, say, 
one percent, an accurate description of such breaking is required (or, at the 
very least, a parametrisation reliable at the one-percent level).
In the light of the quoted PDG error on $V_{us}$, this need becomes 
particularly pressing.

In this talk, after briefly outlining the relation to proton-spin experiments,
I shall discuss the standard SU(3) parametrisation and fit results, together 
with the simple so-called center-of-mass recoil corrections as an attempt to 
account for the major part of the SU(3) breaking.
I shall then move on to describe a new attempt by Xiangdong Ji and myself \cite{Ji97z} at parametrising the violations of SU(3) and conclude the talk 
with a few general observations: an experimental shopping list and a brief 
discussion of the implications for the the $F$ and $D$ parameters and the CKM 
matrix elements.
\section{Motivation}
\subsection{The need for $F$ and $D$}

\vspace{1mm}\noindent
Modulo PQCD corrections (which I shall always suppress but should be understood
as implicitly present) and possible higher-twist contributions (which I shall assume negligible), the nucleon-spin sum rules can be expressed in the 
following form:
\begin{equation}
 \Gamma_1^p \; = \;
 \int_0^1 g_1^p(x) dx \; = \;
 \frac12 \left[ \frac49 \Delta u + \frac19 \Delta d + \frac19 \Delta s \right],
\end{equation}
where it is understood that the polarised quark distributions, $\Delta{q}$,
contain both quark and antiquark contributions and the integral over the momentum fraction, $x$, is also understood.
The analogous sum rule for the neutron is obtained, via an isospin rotation, by 
interchanging $u$ and $d$ in the above expression.
The $\Delta{q}$ appearing in this expression for a high-energy quantity are 
related to the nucleon matrix elements of axial-vector operators:
$\bar\psi_{q'}\frac{\lambda_i}{2}\gamma^3\gamma_5\psi_q$, where the $\lambda_i$ 
are just the relevant SU(3) flavour matrices (diagonal here).
They are therefore also related, via an SU(3) rotation, to precisely the matrix 
elements occurring in the beta decays of the baryon octet, \ie, in HSD.

Thus, the relevant axial coupling constants may be re-expressed in terms of similar sums over polarised quark distributions: \eg,
\begin{equation}
\begin{array}{lcccl}
 \gA^{n\to p}   & = & F+D & = & \Delta u - \Delta d, \\
 \gA^{\SM\to n} & = & F-D & = & \Delta d - \Delta s, \qquad\mbox{\etc}
\end{array}
\end{equation}
There are only the two above linearly-independent non-singlet combinations and 
a third, singlet, constant is supplied by the following sum:
\begin{equation}
 g_0 \; = \; \Delta \Sigma \; = \; \Delta u + \Delta d + \Delta s,
\end{equation}
which, modulo PQCD corrections, clearly measures precisely the total spin 
carried by the quarks inside the nucleon.
This last is then the new constant that is effectively measured in DIS, which 
then allows the separation of the various flavour contributions, {\em providing 
both $F$ and $D$ are known with sufficient accuracy}.

In order to appreciate the importance of the precision in $F$ and $D$, let 
us rewrite the above expressions in order to extract the value of either the
strange-quark spin, $\Delta{s}$, or the total quark contribution, 
$\Delta\Sigma$:
\begin{eqnarray}
 \Delta s     & = & 3\Gamma_1^p - \gA \left(\frac32-\frac{5/3}{1+F/D}\right),\\
 \Delta\Sigma & = & 9\Gamma_1^p - \gA \left(\frac32-\frac{ 1 }{1+F/D}\right),
\end{eqnarray}
where I have taken advantage of the very high precision in the measured value 
of $\gA$ (the experimental error is about 0.2\%).
Thus, independently of the actual results for the nucleon spin measurements one 
sees that (using a typical average value: $F/D\sim0.6$)
\begin{eqnarray}
 \frac{d(\Delta s)}{d(F/D)} & \sim & -2,
\end{eqnarray}
while the sensitivity of $\Delta\Sigma$ is only about half this.
The present nucleon-spin measurements suggest that $\Delta{s}\sim-0.15$, which 
is rather larger than was expected in a na{\"\i}ve quark-parton model picture, 
but which would be completely eliminated by a shift of a mere 0.1 (or 20\%) in 
the value of $F/D$.

\subsection{Extraction of $V_{us}$}

\vspace{1mm}\noindent
Further motivation to improve our understanding of SU(3) breaking in this 
system comes from its possible use for the extraction of $V_{us}$, the CKM 
matrix element.
Angular correlations in the decays are directly given by $\gA/\gV$ while the 
strangeness-conserving (strangeness-changing) decay rates are proportional to 
$|V_{ud}|^2$ ($|V_{us}|^2$).
Thus, at least in principle, an independent measurement of the two elements is 
possible.
However, the rates are also proportional to an overall factor $(\cV\gV^2+\cA\gA^2)$, where the $c_i$ are calculable phase-space factors~\cite{Gar85a}.
Now, while the normalisation of the vector couplings is well protected by the
Ademollo-Gatto theorem \cite{Ade64a} (which states that any breaking effects 
are second order here), there is no such control over the axial couplings.
Indeed, as we shall see from the following discussion, the breaking effects are
found experimentally to be typically of the order of a few percent.

The error on $V_{us}$ quoted by the PDG above implies a knowledge of $\gA$ to
better than 1\%.
In contrast, while the value of $V_{us}$ turns out to be rather stable with
respect to SU(3) breaking in the axial couplings, the values one extracts 
from present data with the method that best describes them unfortunately 
implies a violation of CKM unitarity of the order of 2\%.
In order to properly account for this, a specific model for the breaking is 
required. 
However, the only such calculation that might allow extraction of the couplings 
does not, as we shall see, describe the data at all successfully.
\section{The Data}

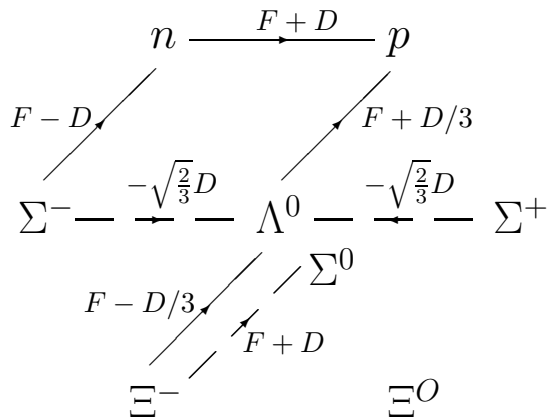
\begin{figure}[hbt]
\begin{center}
\begin{picture}(20000,16000)
 \put( 6000,15000){\Large$n$}
 \put(15000,15000){\Large$p$}
 \put( 1000, 8000){\Large$\SM$}
 \put(10000, 8000){\Large$\LO$}
 \put(12000, 6000){\Large$\SO$}
 \put(19000, 8000){\Large$\SP$}
 \put( 5200, 1000){\Large$\XM$}
 \put(15000, 1000){\Large$\XO$}
 \drawline\fermion[\E \REG]( 7500,15300)[7000]
 \drawarrow[\E\ATTIP](11300,15300)
 \put(10000,15700){\small$F+D$}
 \drawline\scalar [\E \REG]( 3200, 8500)[3]
 \drawarrow[\E\ATTIP]( 6500, 8500)
 \put( 5100, 9500){\small$-\sqrt\frac23D$}
 \drawline\scalar [\W \REG](18200, 8500)[3]
 \drawarrow[\W\ATTIP](15000, 8500)
 \put(14100, 9500){\small$-\sqrt\frac23D$}
 \drawline\fermion[\NE\REG]( 2000,10000)[6000]
 \drawarrow[\NE\ATTIP]( 4300,12300)
 \put(  700,12000){\small$F-D$}
 \drawline\fermion[\NE\REG](11000,10000)[5800]
 \drawarrow[\NE\ATTIP](13300,12300)
 \put(14000,12000){\small$F+D/3$}
 \drawline\fermion[\NE\REG]( 6000, 3000)[6000]
 \drawarrow[\NE\ATTIP]( 8300, 5300)
 \put( 3500, 5000){\small$F-D/3$}
 \drawline\scalar [\NE\REG]( 7500, 2500)[3]
 \drawarrow[\NE\ATTIP]( 9800, 4800)
 \put( 9500, 3500){\small$F+D$}
\end{picture}
\caption{The measured HSD's: the solid lines represent those decays where both the rate and angular correlations are measured and the dashed lines, those for 
which only rates are known.}
\label{fig:decs}
\end{center}
\end{figure}

\vspace{1mm}\noindent
Before going on to fit results and corrections, we should briefly examine the 
data and, in particular, we have to deal with the long-standing problem of the discrepancies presented by the neutron beta-decay data alone.
The measured decay channels are represented schematically in 
fig.~\ref{fig:decs}.

The values for the rates and angular correlations are given together with the
relation to $F$ and $D$ in table~\ref{tab:data}.
\begin{table}[hbt]
\caption{The hyperon semi-leptonic decay data, as used in this 
analysis~\protect\cite{PDG96a}. 
The figures marked in bold type-face are those measured to a precision of 
better than 5\%.} 
\vspace*{3pt}
\centering
\boxnotes{$\begin{array}
 {|l@{\,\to\,}l|l@{\,\pm\,}l|l@{\,\pm\,}l|l@{\,\pm\,}l|l|c|} \hline
 \multicolumn{1}{|c }{} & & 
 \multicolumn{4}{ c|}{\mbox{Rate ($10^6\,$s$^{-1}$)}}          &
 \multicolumn{2}{ c|}{\mbox{Angular correlations}}	     & &
 \\ \cline{3-6}
 \multicolumn{2}{|c|}{\raisebox{1.5ex}[0pt][0pt]{Decay}}       & 
 \multicolumn{2}{ c|}{\ell=e}                                  &
 \multicolumn{2}{ c|}{\ell=\mu}                                & 
 \multicolumn{2}{ c|}{\raisebox{0.0ex}[0pt][0pt]{$\gA/\gV$}}   &
 \multicolumn{1}{ c|}{\raisebox{1.5ex}[0pt][0pt]{$\gA/\gV$}}   &
 \multicolumn{1}{ c|}{\raisebox{1.3ex}[0pt][0pt]{$|\Delta S|$}}
 \\ \hline  n     &  p\,\ell\bar\nu & \mathbf 1.1274  & \mathbf 0.0025
\ftnt{Rate in $10^{-3}\,$s$^{-1}$.}                        &
 \NoCol &         &\mathbf \PM1.2599        &\mathbf 0.0025
\ftnt{Taken from ref.~\protect\cite{Tow95a}, using the PDG value of 
 $1.2601\pm0.0025$ makes a negligible difference.}
                                              & F+D        & 0
 \\ \hline \LO    &  p\,\ell\bar\nu &\mathbf 3.161   &\mathbf 0.058      &
 0.60   & 0.13    &\mathbf\PM0.718         &\mathbf 0.015   & F+D/3      & 1
 \\ \hline \SM    &  n\,\ell\bar\nu &\mathbf 6.88    &\mathbf 0.23       &
 3.04   & 0.27    &\mathbf  -0.340         &\mathbf 0.017   & F-D        & 1
 \\ \hline \SM    &  \LO\ell\bar\nu &\mathbf 0.387   &\mathbf 0.018      & 
 \NoCol &         &  \NoCol         &         & -\sqrt{\frac23}\,D
\ftnt{$\gV=0$, absolute expression for $\gA$ given.}
							   & 0
 \\ \hline \SP    &  \LO\ell    \nu & 0.250   & 0.063      &
 \NoCol &         &  \NoCol         &         & -\sqrt{\frac23}\,D\,^c
							   & 0
 \\ \hline \XM    &  \LO\ell\bar\nu & 3.35    & 0.37
\ftnt{Error scale factor of 2 included, following the PDG practice for
 discrepant data.}
                                                           &
 2.1    & 2.1
\ftnt{Not used in fits.}
                  &\PM0.25          & 0.05    & F-D/3      & 1
 \\ \hline \XM    &  \SO\ell\bar\nu & 0.53    & 0.10       &
 \NoCol &         &  \NoCol         &         & F+D        & 1
 \\ \hline 
\end{array}$}
\label{tab:data}
\end{table}
Note that a large fraction of the data are known to a precision of 5\% or 
better and thus dealing with the problem of SU(3) breaking is clearly highly 
desirable at this point.
One can perform various internal consistency checks, \eg, between the 
individual decay values from rates and angular correlations.
To the precision required there is no outstanding problem; however, there is a
notable difficulty in the best measured sector: that of the neutron.

The beta decay of the neutron is now measured to a precision of two parts per
thousand (in both the rate and the angular correlations).
This highlights a discrepancy, which, while being very small in absolute terms, 
is rather large in terms of the number of standard deviations.
Thus, in order not to cloud the examination of HSD with an artificially large 
overall $\chi^2$, I have decided to use weighted averages and then rescale the 
errors involved with the resulting $\sqrt{\chi^2}$.
This in no way biases the HSD fits, especially since the variations of the 
central values are typically less than one part per thousand, and the spurious 
contribution to the overall $\chi^2$ is completely eliminated.
For more details, the interested reader is referred to 
refs.~\cite{Rat90b,Rat95b}.
\section{Fitting the Data} \label{sec:fits}

\subsection{Na{\"\i}ve SU(3) fits}

\vspace{1mm}\noindent
The simplest possible fit assumes SU(3) symmetry and uses three free 
parameters: $F$, $D$ and $V_{ud}$ (constraining $V_{us}$ by CKM unitarity).
The results of such a fit, in which no external value of $V_{ud}$ is input, is
shown in table~\ref{tab:su3fita}, where different data subsets are also
compared.
\begin{table}[hbt]
\caption{An SU(3)-symmetric fit to the data.}
\vspace*{3pt}
\centering
\boxnotes{$\begin{array}
 {|c|l@{\,\pm\,}l|l@{\,\pm\,}l|l@{\,\pm\,}l|c|c|} \hline
 & \multicolumn{6}{c|}{\mbox{Parameters}} & &
 \\ \cline{2-7}
 \multicolumn{1}{|c|}{\raisebox{1.5ex}[0pt][0pt]{Data}}         & 
 \multicolumn{2}{ c|}{V_{ud}}                                   &
 \multicolumn{2}{ c|}{F}                                        & 
 \multicolumn{2}{ c|}{D}                                        & 
 \multicolumn{1}{ c|}{\raisebox{1.5ex}[0pt][0pt]{$\chi^2$/DoF}} &
 \multicolumn{1}{ c|}{\raisebox{1.5ex}[0pt][0pt]{$F/D$}}
 \\ \hline
 \mbox{Rates}      & 0.9749 & 0.0004 & 0.469 & 0.008 & 0.797 & 0.008
                   & 3.8    & 0.589 \\ \hline 
 \mbox{Ang. Corr.} &
 \multicolumn{2}{c|}{-\rlap{\ftnt{Undetermined in this fit.}}}
                                     & 0.460 & 0.008 & 0.800 & 0.008
                   & 0.8    & 0.576
 \\ \hline 
 \PV\Delta S\PV=0  & 0.9795 & 0.0020 & 0.528 & 0.017 & 0.732 & 0.017
                   &  -\rlap{\ftnt{Zero degrees of freedom.}}
                            & 0.721
 \\ \hline 
  |\Delta S|   =1  & 0.9742 & 0.0006 & 0.448 & 0.009 & 0.791 & 0.017
                   & 0.8    & 0.567
 \\ \hline 
 \mbox{All}        & 0.9750 & 0.0004 & 0.465 & 0.006 & 0.799 & 0.006
                   & 3.0    & 0.582
 \\ \hline 
 \mbox{All}+V_{ud}
\ftnt{$V_{ud}$ taken from nuclear $ft$ and $\Kl3$ analysis.}
                   & 0.9751 & 0.0002 & 0.465 & 0.006 & 0.799 & 0.006
                   & 2.7    & 0.583
 \\ \hline 
\end{array}$}
\label{tab:su3fita}
\end{table}
Let me draw your attention to the perfectly acceptable $\chi^2$ returned by the 
subset of angular correlation data only.
This means, in particular, that {\em the study of the angular correlation data
alone is of absolutely no relevance to the problem of SU(3) breaking}, a fact
that is overlooked in many papers, even those purporting to find surprisingly
large effects.

Since the main purpose here is the study of SU(3) breaking on $F$ and $D$, it 
is then convenient to use the external knowledge of $V_{ud}$ supplied by the 
$\Kl3$ and also the $ft$ values in the so-called super-allowed $0^+$-$0^+$ 
nuclear beta transitions.
The results of such a series of fits are presented in table~\ref{tab:su3fitb}.
\begin{table}[hbt]
\caption{An SU(3)-symmetric fit to the modified data including the external 
$V_{ud}$ from nuclear $ft$ and $\Kl3$ analysis (see text for details).}
\vspace*{3pt}
\centering
\boxnotes{$\begin{array}
 {|c|l@{\,\pm\,}l|l@{\,\pm\,}l|l@{\,\pm\,}l|c|c|} \hline
 & \multicolumn{6}{c|}{\mbox{Parameters}} & &
 \\ \cline{2-7}
 \multicolumn{1}{|c|}{\raisebox{1.5ex}[0pt][0pt]{Data}}         & 
 \multicolumn{2}{ c|}{V_{ud}}                                   &
 \multicolumn{2}{ c|}{F}                                        & 
 \multicolumn{2}{ c|}{D}                                        & 
 \multicolumn{1}{ c|}{\raisebox{1.5ex}[0pt][0pt]{$\chi^2$/DoF}} &
 \multicolumn{1}{ c|}{\raisebox{1.5ex}[0pt][0pt]{$F/D$}}
 \\ \hline
 \mbox{Rates}      & 0.9749 & 0.0003 & 0.469 & 0.008 & 0.796 & 0.009
                   & 3.2    & 0.589
 \\ \hline 
 \mbox{Ang. Corr.} & 0.9752 & 0.0007 & 0.460 & 0.008 & 0.799 & 0.009
                   & 0.8    & 0.576
 \\ \hline 
 \PV\Delta S\PV=0  & 0.9753 & 0.0007 & 0.529 & 0.017 & 0.735 & 0.017
                   & 0.5    & 0.719
 \\ \hline
  |\Delta S|   =1  & 0.9747 & 0.0005 & 0.452 & 0.009 & 0.799 & 0.015
                   & 0.8    & 0.566
 \\ \hline 
 \mbox{All}        & 0.9749 & 0.0003 & 0.465 & 0.006 & 0.798 & 0.006
                   & 2.3    & 0.582
 \\ \hline 
\end{array}$}
\label{tab:su3fitb}
\end{table}
The main point to be stressed is that very little changes and that the overall 
$\chi^2$ improves, since the values of $V_{ud}$ almost coincide.
Note also that the value of $F/D$ is very stable indeed.

Before moving on to the breaking fits, I should make a brief comment on the 
situation regarding $V_{us}$.
If one now attempts to extract $V_{us}$ independently (\ie, not by invoking CKM 
unitarity), then one obtains
\begin{eqnarray}
 V_{us} & = & 0.223 \pm 0.002.
\end{eqnarray}
Such a value is marginally in violation of CKM unitarity (in excess).

I should also comment on the glaring and related discrepancy between $F/D$ values extracted from the strangeness-changing and strangeness-conserving 
decays.
Even taking into account the larger errors present for the two subsets, the 
difference is still in the order of several standard deviations. 
The fact that the dependence on $F$ and $D$ in these two subsets is rather 
different should warn us to be very wary that this discrepancy could be 
driven mainly by the very flavour violations under study.

\subsection{SU(3) breaking fits}

\vspace{1mm}\noindent
One of the most successful models for describing SU(3) breaking in this sector 
is based on the so-called centre-of-mass (CoM) or recoil 
correction~\cite{Don87a}.
The idea is simply to describe the hadron as an extended object by taking into
account the smearing of the wave function due to non-zero momentum.
Calculation of the relevant matrix elements then leads to correction 
formul{\ae} that depend on just one parameter $\langle{p^2}\rangle$, the mean 
three-momentum squared in the wave function, and on the masses of the baryons 
involved in a given decay.
To calculate this parameter the authors of \cite{Don87a} adopted a bag-model
approach, obtaining $\langle{p^2}\rangle\simeq0.43\,$GeV$^2$. 
Such an approach also provided a means to calculate the corrections arising 
from a possible strange-quark wave-function overlap mismatch.
While it turns out that the former correction parameter actually corresponds to
its minimum-$\chi^2$ value, the latter is far too large (it is about 8\% for 
the strangeness-changing decays) and actually worsens the fit 
considerably~\cite{Rat90b,Rat95b}.

The results of this simple CoM SU(3)-breaking fit (\ie, without the 
wave-function correction) are shown in table~\ref{tab:comfit}.
\begin{table}[hbt]
\caption{An SU(3)-breaking fit to the modified data including the external, 
world average, value for $V_{ud}$.} 
\vspace*{3pt}
\centering
\boxnotes{$\begin{array}
 {|c|l@{\,\pm\,}l|l@{\,\pm\,}l|l@{\,\pm\,}l|c|c|} \hline
 & \multicolumn{6}{c|}{\mbox{Parameters}} & &
\\ \cline{2-7}
 \multicolumn{1}{|c|}{\raisebox{1.5ex}[0pt][0pt]{Data}}         & 
 \multicolumn{2}{ c|}{V_{ud}}                                   &
 \multicolumn{2}{ c|}{F}                                        & 
 \multicolumn{2}{ c|}{D}                                        & 
 \multicolumn{1}{ c|}{\raisebox{1.5ex}[0pt][0pt]{$\chi^2$/DoF}} &
 \multicolumn{1}{ c|}{\raisebox{1.5ex}[0pt][0pt]{$F/D$}}
 \\ \hline
 \PV\Delta S\PV=0  & 0.9753 & 0.0007 & 0.481 & 0.018 & 0.784 & 0.018
                   & 0.5    & 0.613 \\ \hline
  |\Delta S|   =1  & 0.9747 & 0.0005 & 0.465 & 0.009 & 0.825 & 0.015
                   & 1.0    & 0.563 \\ \hline 
 \mbox{All}        & 0.9744 & 0.0003 & 0.460 & 0.006 & 0.806 & 0.006
                   & 1.0    & 0.571 \\ \hline 
\end{array}$}
\label{tab:comfit}
\end{table}
From a detailed examination of such fits, it emerges that by far the largest 
effects are found in the $\SC\to\LO$ decays, but this might have been expected 
as they have no $\gV$ contribution.
In any case, the overall distribution of the deviations in the final fit is
perfectly normal; and so no statement can be made as to the unreliability or
otherwise of any particular data point.
As far as the possibility of extracting $V_{us}$ separately is concerned, it
should be noted that the fits are improved somewhat by allowing a 
renormalisation of the strangeness-changing decay rates.
The factor required is a shift of about 2\% (at the $1.5\sigma$ level) in the
direction of reducing the net overshoot of CKM unitarity.

Moreover, returning to the earlier discrepancy between the $|\Delta{S}|=0$ and 
1 decays, we see now that it has indeed entirely disappeared.
On the one hand, this comforts us that the model is working well; but, on the 
other, we should take this as a further strong indication of the sensitivity of
$V_{us}$ to SU(3) breaking, which I stress is not at all under control with 
respect to this particular aspect of the decays.
On the contrary, this aspect of the decays represents the only handle on the effects of the breaking between the $|\Delta{S}|=0$ and 1 decays.
\section{Towards a Minimal ``Model-Independent'' Description}

\vspace{1mm}\noindent
A recent approach, where generality was the key note, attempted to investigate
all possible SU(3)-breaking structures~\cite{Song}.
The idea was that the breaking should be driven by an effective interaction 
with the SU(3) structure of $\lambda_8$, the hypercharge matrix.
One can show that, for the decays in question, there are four independent 
structures that can contribute.
Unfortunately, the quality and quantity of the present data do not allow the 
determination of as many as four extra parameters and so the effect of each one 
can only be tried separately.
However, since the best form could be any arbitrary combination, fitting in 
this fashion is of little use in attempting to understand the true breaking 
pattern.

In an attempt to derive a general form for the breaking pattern and, in 
particular, to justify the success of models based on mass-driven breaking, 
Xiangdong Ji and I \cite{Ji97z} have recently tried to formulate a breaking 
scheme based directly on the well-known and very successful Gell-Mann--Okubo 
mass formul{\ae}~\cite{GMO}.
The idea then is to assume that the breaking is due solely to effects of the 
strange-quark mass and thus we separate the strong hamiltonian into its SU(3) 
symmetric and violating pieces: ${\cal H}={\cal H}_0+{\cal H}'$ and take 
${\cal H}'$ as transforming as $\lambda_8$.
Thus, as is well-known, the mass formul{\ae} are then 
\begin{eqnarray}
 m_B & = & m_0 + m_a \tr(\lambda_8 B \bar{B}) + m_b \tr(\lambda_8 \bar{B} B),
\end{eqnarray}
where $m_a$ and $m_b$ are two small mass parameters, which indeed allow the octet masses to be described very accurately.

If we now ascribe SU(3) breaking in HSD to the same interaction operator 
(${\cal H}'$) then, since the breaking is known to be small, we can adopt the 
following first-order perturbation-series approximation for the baryon states:
\begin{eqnarray}
 \ket{B} & = & \ket{B}_0
         + \sum_{N\not=B} \ket{N} \frac{\bra{N}{\cal H}'\ket{B}_0}{(m_0-m_N)},
\label{eq:states}
\end{eqnarray}
where the (unperturbed) states, $\ket{N}$, have the same flavour (and, by 
hypothesis, spin) quantum numbers as $\ket{B}_0$ (since ${\cal H}'$ is 
diagonal) and simply belong to a higher-mass representation.

Now, since the physical $\ket{B}$ of interest here are actually the 
lowest-lying states we can make our first serious assumption: namely, that the 
next-lowest-lying state $\ket{N}$ dominates the sum---it has the smallest 
energy denominator and is also likely to have the largest wave-function overlap 
with $\ket{B}_0$.
Thus, the sum may be approximated by
\begin{eqnarray}
 \ket{B} & \simeq & \ket{B}_0
         - \frac{1}{\Delta} \sum_{N\not=B} \ket{N}\bra{N}{\cal H}'\ket{B}_0,
\end{eqnarray}
where $\Delta=m^{\mathrm{lowest}}_N-m_0$ is a constant, equal for all baryons 
of the octet.
Adding and subtracting the missing (first) term in the (otherwise complete) 
sum, one then obtains, in the so-called closure approximation
\begin{eqnarray}
 \ket{B} & \simeq & \left( 1+\frac{\delta m}{\Delta} \right) \ket{B}_0
                  - \frac{1}{\Delta} {\cal H}' \ket{B}_0 \nonumber \\
         &    =   & \left[ 1+\frac{1}{\Delta}
                           \left(\delta m-{\cal H}'\right) \right] \ket{B}_0,
\end{eqnarray}
where $\delta{m}=\prefix_0{\bra{B}}{\cal{H}}'\ket{B}_0=m-m_0$ gives just the 
mass shifts with respect to some common central reference point.
Finally, inserting this expression into the usual expressions for the 
axial-vector matrix elements, one obtains
\begin{eqnarray}
 g_A & = & \left( 1 + \frac{\delta m_i+\delta m_f}{\Delta} \right)
       \prefix_0{\bra{B_f}} A_W^3 \ket{B_i}_0
     - \frac{1}{\Delta} \prefix_0{\bra{B_f}} \{A_W^3,{\cal H}'\} \ket{B_i}_0,
\label{eq:avme}
\end{eqnarray}
where $i,f$ stand for the initial and final states and $A_W^3$ is just the 
third component of the relevant weak axial-vector operator.

The flavour matrix structure of the last, inhomogeneous, term now turns out to 
be rather simple, since the relevant {\em matrices\/} satisfy the following 
simple anticommutation relations, which would maintain the flavour structure 
unaltered:
\begin{equation}
\begin{array}{rcrl}
 \{\lambda^\pm_I,\lambda_8\}
 & = &   {\textstyle\frac{2}{\sqrt3}} \lambda^\pm_I 
 & \qquad (\lambda^\pm_I = \lambda_1 \pm i\lambda_2),
\\
 \{\lambda^\pm_V,\lambda_8\}
 & = & - {\textstyle\frac{2}{\sqrt3}} \lambda^\pm_V
 & \qquad (\lambda^\pm_V = \lambda_4 \pm i\lambda_5).
\end{array}
\end{equation}
However, the anticommutators of the {\em operators\/} concerned do not simplify 
in the usual way and thus this term does not reduce to the same structure as 
leading term, as we had hoped.
What remains as an extra, undesirable, piece is a sort of diquark-diquark 
correlator.
Therefore, if we assume, as is likely, that such an object is rather 
suppressed, then the corrections to $g_A$ may simply be written in the 
following form: 
\begin{eqnarray}
 g_A & = & g_A^0 \, \left( 1 + \frac{\delta m_i+\delta m_f}{\Delta}
                           \pm \epsilon \right),
\end{eqnarray}
where $\epsilon$ is one new free parameter describing the SU(3) breaking 
between strangeness-conserving ($+$) and strangeness-changing ($-$) decays 
(\ie, the imperfect strange and non-strange quark wave-function overlap).
The term in $\epsilon$ was already effectively considered above \cite{Rat95b}, 
but found to be small and not well determined, and thus should be neglected at 
this level of analysis (it is, in fact, of the same order of magnitude as the 
neglected higher-order terms).

Clearly, the failure to reduce the expressions to the final form cleanly (\ie, 
without further assumptions) weakens the arguments for the simplicity of the 
breaking structure.
One could, however, hope to salvage some of the predictiveness of this approach 
by taking a step back to the expression for the SU(3) broken states, 
eq.~(\ref{eq:states}).
The intermediate states in eq.~(\ref{eq:states}), $\ket{N}\bra{N}$, can be 
decomposed in terms of the possible irreducible representations contained in 
the product:
\begin{eqnarray}
  8\otimes8 & = & 1\oplus8\oplus8\oplus10\oplus\overline{10}\oplus27.
\end{eqnarray}
Selecting just one class (\eg, the 10, which actually goes hand-in-hand with 
the $\overline{10}$ owing to the symmetric nature of the correction terms, \cf, 
eq.~\ref{eq:avme}), one could then compare the breaking pattern so determined 
(obtained by fitting the data) with that obtained from any given model.
Such a comparison should provide clear indications as to the validity or 
otherwise of the model.
\section{Conclusions}

\subsection{Experimental shopping list}

\vspace{1mm}\noindent
From an examination of the experimentally observed pattern of breaking and the 
sensitivity of the fits to the various input data, one can make the case for a 
more detailed experimental study of the following decays:
\begin{itemize}
\item 
$\SC\to\LO\ell\nu$: not only is the breaking seen to be largest here, but this 
also represents the only $\Delta{S}=0$ decay besides that of the neutron, 
whence the presence of an external, independent, value for $V_{ud}$ would then 
allow a very direct examination of SU(3) breaking;
\item 
$\SM\to{ne}\bar\nu$ and $\LO\to pe\bar\nu$: these two decays can provide a 
better indicate as to the presence or otherwise of so-called second-class 
currents, which might be large here and which, it has been noted, would tend to 
increase the extracted value of $F/D$~\cite{Hsu88a};
\item 
$\XM\to\SO{e}\bar\nu$: this is the decay where the corrections might be 
expected to be largest of all, it also has the merit of depending on $F+D$, 
which is known to a very high precision, and could therefore act as a solid 
point of reference for breaking effects.
\end{itemize}

\subsection{Theoretical considerations}

\vspace{1mm}\noindent
In conclusion then, let me make the following observations on the HSD 
phenomenology that has been discussed here:
\begin{itemize}
\item 
the CoM corrections are physically motivated and provide a perfectly acceptable 
description of SU(3) breaking in this sector, neither needing nor indeed 
leaving room for any additional corrections (barring a minor global 
renormalisation of the strangeness-changing decay amplitudes);
\item  
a rather more general (though admittedly not entirely model-independent) 
purely {\em mass\/}-driven breaking also performs acceptably well.
\item 
it makes absolutely no sense to fit the present angular-correlation data 
{\em alone\/} since the breaking is completely negligible there within 
experimental errors---indeed, there is strong risk of susceptibility to 
statistical fluctuations;
\item
$V_{us}$ extracted from HSD {\em overshoots\/} CKM unitarity and therefore
demands a further $\sim2\%$ renormalisation of the $|\Delta{S}|=1$ decays, thus 
the error quoted by the PDG would appear to make no sense at all;
\item
the present best fit gives $F/D=0.57\pm0.01$ \cite{Rat95b} and therefore the 
indications for a non-zero strange-quark polarisation cannot easily be 
eliminated by merely appealing to SU(3)-breaking effects.
\end{itemize}
\section{Acknowledgments}

\vspace{1mm}\noindent
The author is very happy to acknowledge stimulating conversations with Profs.\ 
Tom Cohen and Xiaotong Song, and the hospitality of both the Institute for 
Nuclear Theory at the University of Washington, Seattle and the University 
of Maryland, where some of the work presented was carried out.

\end{document}